\documentclass[a4paper]{jpconf}
\usepackage{graphicx}
\begin{document}
\title{EDELWEISS-II Dark Matter Search : status and first results}

\author{V\'eronique SANGLARD for the EDELWEISS collaboration}

\address{IPNL, Universit\'e de Lyon, Universit\'e Lyon 1, CNRS/IN2P3, 4 rue Enrico Fermi, 69622 Villeurbanne, France}

\ead{sanglard@ipnl.in2p3.fr}

\begin{abstract}
The EDELWEISS II experiment is devoted to the search
for Weakly Interacting Massive Particles (WIMP) that would
constitute the Dark Matter halo of our Galaxy.
For this purpose, the experiment uses cryogenic
germanium detectors, cooled down at 20~mK, in which the
collision of a WIMP with an atomic nucleus produces characteristic
signals in terms of ionization and elevation of temperature.
We will present the preliminary results of the first
operation of the detectors installed in the underground
laboratory of the Frejus Tunnel (LSM), attesting to the
very low radioactive background conditions achieved so far.
New detectors, with a special electrode design for
active rejection of surface events, have been shown to be
suited for searches of WIMPs with scattering
cross-sections on nucleon well below 10$^{-8}$~pb.
Preliminary results of WIMP search performed
with a first set of these detectors are presented.
\end{abstract}

\section{Introduction}
Recent cosmological observations of the CMB show that most ($\sim$~85$\%$) of the 
matter in our Universe is dark and non baryonic. If non baryonic Dark Matter is made of particles,
they must be stable, neutral and massive : WIMPs (Weakly Interactive Masssive Particles). 
In the MSSM (Minimal Supersymmetric Standard Model) framework, the WIMP could be the LSP 
(Lightest Supersymmetric Particle) called neutralino. Its mass is expected to lie between few tens and 
few hundreds of GeV/c$^2$, and a scattering cross section with a nucleon below $10^{-6}$~pb.\\  
The EDELWEISS experiment is dedicated to the direct detection of WIMPs. 
The direct detection method (used also by other experiments like CDMS~\cite{cdms1}, XENON~\cite{xenon} and CRESST~\cite{cresst}) consists in the measurement of the energy released by nuclear recoils produced by the elastic collision of a WIMP from the galactic halo with an atomic nucleus.
The main challenge is the expected extremely low event rate ($\leq$ 1~evt/kg/year) due to the very small interaction cross-section of WIMP with  nucleons. The ability of the detector to unambiguously tag nuclear recoil interactions plays a major role in these searches.  An other constraint is the relatively small deposited energy  (a few keV to a few tens of keV). 
\section{Experimental set-up}
\begin{figure}[h]
\begin{center}
\includegraphics*[width=15pc]{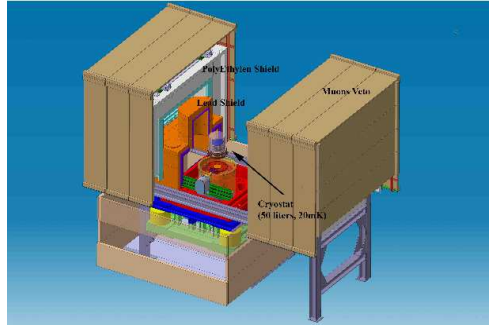}
\caption{General scheme of the EDELWEISS-II experiment.}
\label{fig:setup}
\end{center}
\end{figure}

The EDELWEISS experiment is located in the Modane Underground Laboratory (LSM) in the Fr\'ejus tunnel 
connecting France and Italy under $\sim$~1700~m of rock ($\sim$~4800~mwe). 
In the laboratory, the muon flux is reduced down to 4~$\mu$/m$^2$/d and the fast neutron flux has been 
measured to be $\sim$~1.6$\times$ 10$^{-6}$~cm$^2$/s~\cite{neutron}.
The EDELWEISS-II experiment installation (see Fig.~\ref{fig:setup}) was completed end of 2005. Specific improvements have been made in order to reduce the possible background sources that have limited the sensitivity of the previous phase EDELWEISS-I~\cite{{simon},{edel6}}
To reduce the radioactive background in the cryostat all the materials were tested for 
radiopurity in a HPGe dedicated detector with very low radon level. The experiment is installed in a class 10 000 clean room  
and the cryostat environment is submitted to a permanent flow of deradonized air. 
The gamma background is screened by a 20~cm thick lead shield.
Concerning the low energy neutron background, due to the radioactive surrounding rock, it is attenuated 
by more than three orders of magnitude thanks to a 50~cm polyethylene shielding. 
In addition, a muon veto surrounding the experiment will tag muons interacting
in the lead shielding~\cite{veto}.  The dilution cryostat is of inverted design, with the experimental detector volume on the top. The large volume, 50~$\ell$, allows running 120 identical detectors in a compact arrangement, that will improve the possibility of 
detecting multiple interactions of neutrons and hence reject them. 
The detectors used in the experiment are cryogenic bolometers with simultaneous measurement of
phonon and ionization signals, cooled at a temperature of $\sim$20~mK. They are made of a cylindrical high purity Ge crystal with Al 
electrodes to collect ionization signals and a NTD heat sensor glued onto one electrode
to collect the phonon signal~\cite{edel3}. 
The simultaneous measurement of both heat and ionization signals provides an excellent 
event by event discrimination between nuclear recoils (induced by WIMP or neutron 
scattering) and electron recoils  (induced by $\alpha$, $\beta$ or $\gamma$-radioactivity). The ratio
of the ionization and heat signals depends on the recoiling particle, since a nucleus
produces about one third of the ionization in a crystal than an electron does.

\section{Rejection of surface event with ID detectors}
The EDELWEISS experiment is currently running in its second phase. The first, EDELWEISS-I~\cite{edel6}, has been limited by surface events generated by the $\beta$ radioactivity of $^{210}$Pb~\cite{simon}. Because of diffusion, trapping and recombination, the charge induced by surface events is miscollected and can mimic nuclear recoils. Thus, they cannot be actively rejected with the standard EDELWEISS detectors. 
\begin{figure}[h]
\begin{center}
\includegraphics[width=14pc]{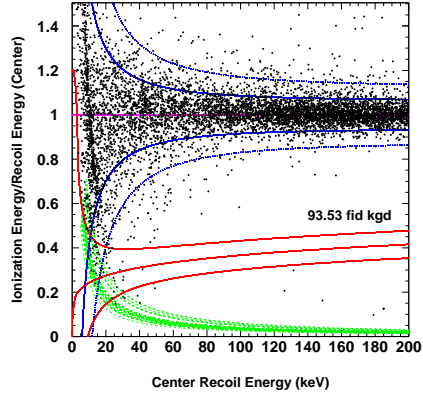}
\caption{\label{run8} Ionization yield versus recoil energy for a 93.5~kg.d exposure using 11 standard EDELWEISS detectors. The leaking events down to the nuclear recoil band are attributed to $\beta$ interactions.}
\end{center}
\end{figure}
This may be seen in the data taken in 2008 with 11 detectors. An exposure of 93.5~kg.d was achieved with an analysis threshold set {\it a priori} to 30~keV. The $\alpha$ and $\beta$ backgrounds were reduced in comparison to the EDELWEISS-I ones but, as it can be seen in the figure~\ref{run8}, it is still the limiting factor to improve sensitivity.
To solve the problem of surface events, new detectors with active surface rejection were studied during last years inside the EDELWEISS facility. These new detectors, named InterDigits (ID) have a mass of the order of 400~g. The ionization electrodes are replaced by two sets of concentric, annular interleaved electrodes polarized differently~\cite{{id1},{id2}}. Surface events are tagged by the presence of charge on two electrodes on the same side of the detector.\begin{figure}[h]
\begin{minipage}{17pc}
\includegraphics[width=16pc]{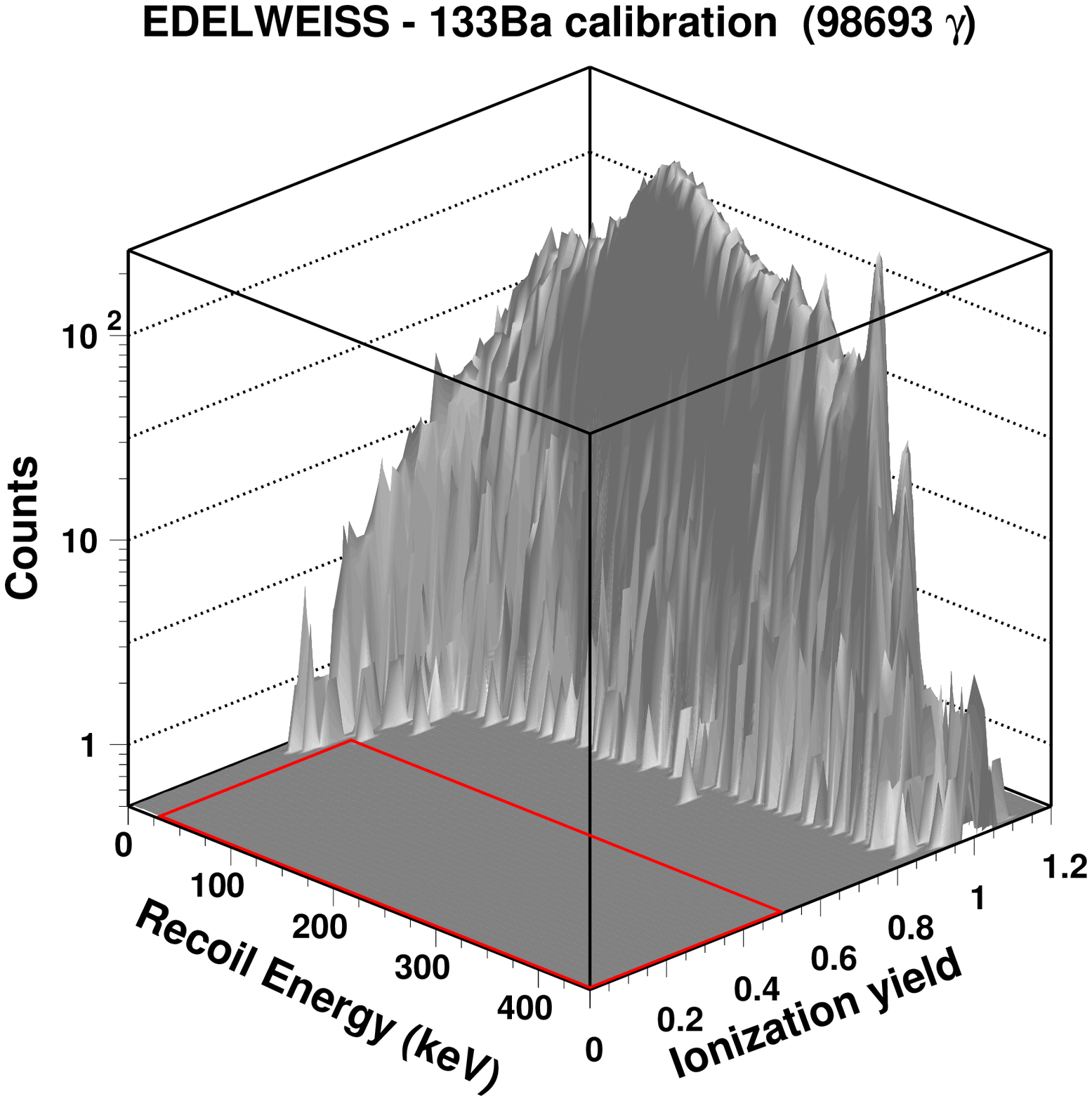}
\caption{\label{id1} ID detectors calibration with a $^{133}$Ba source.}
\end{minipage}\hspace{4pc}%
\begin{minipage}{17pc}
\begin{center}
\includegraphics[width=16pc]{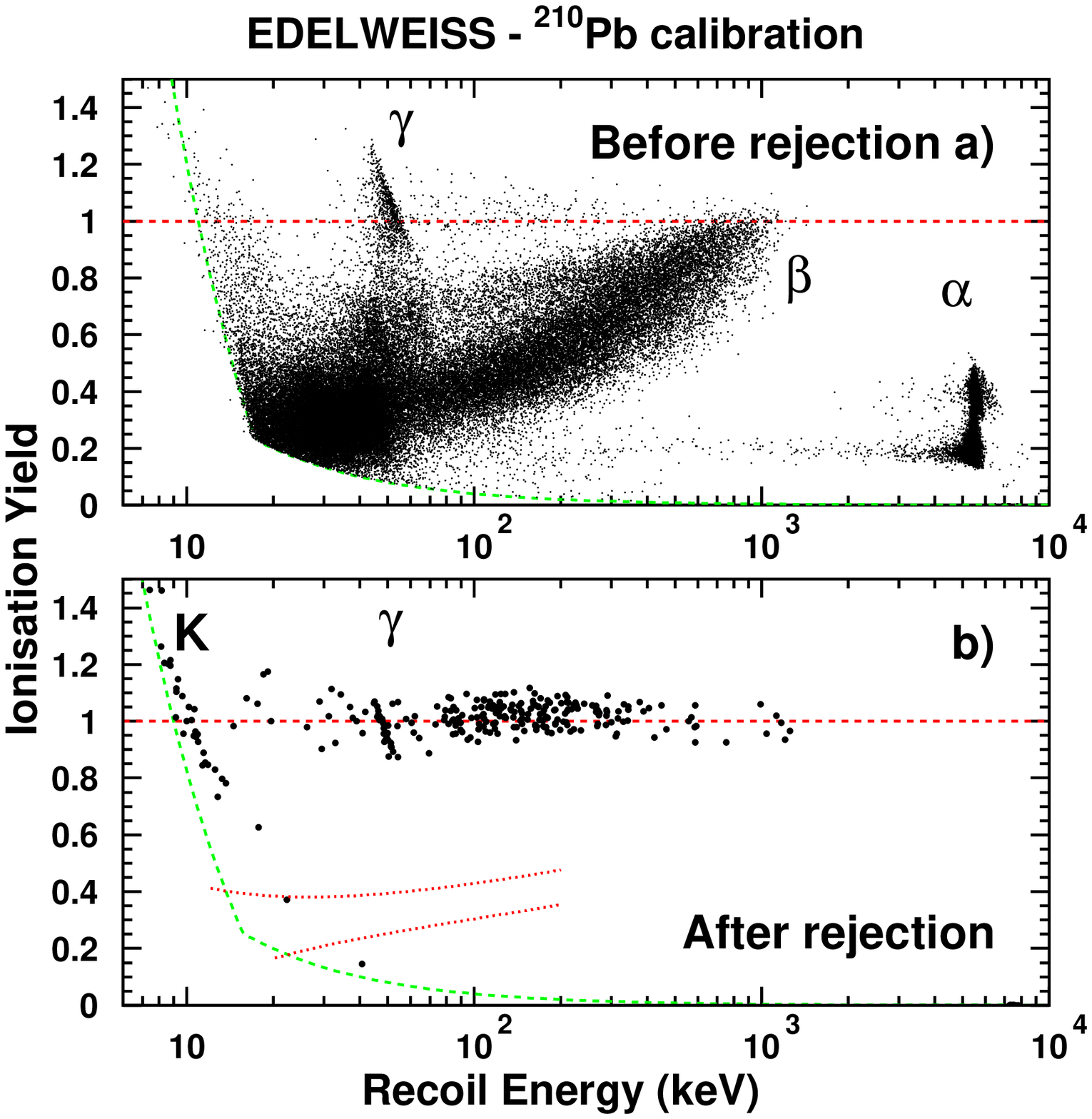}
\caption{\label{id2} ID detectors calibration with a $^{210}$Pb source.}
\end{center}
\end{minipage} 
\end{figure}
As shown in figures~\ref{id1} and~\ref{id2}, this method is very efficient to reject surface events. Figure~\ref{id1} shows the ionization yield as a function of recoil energy for high-statistics calibration performed with a $^{133}$Ba source. No events are present after the fiducial selection in the nuclear recoil band resulting in a rejection factor better than 1 in 10$^5$ for $\gamma$ rays below 60~keV. Figure~\ref{id2} shows the same kind of plot for a calibration with a $^{210}$Pb source. Before the rejection of surface events, we see a lot of events in the nuclear recoil band but only one after giving a rejection factor of about 1 in 10$^5$. Such performances are described in details in~\cite{id3} and open the way to reach sensitivities below 10$^{-8}$~pb.  

\section{First results of the ID detectors}
\begin{figure}[h]
\begin{minipage}{17pc}
\includegraphics[width=15pc]{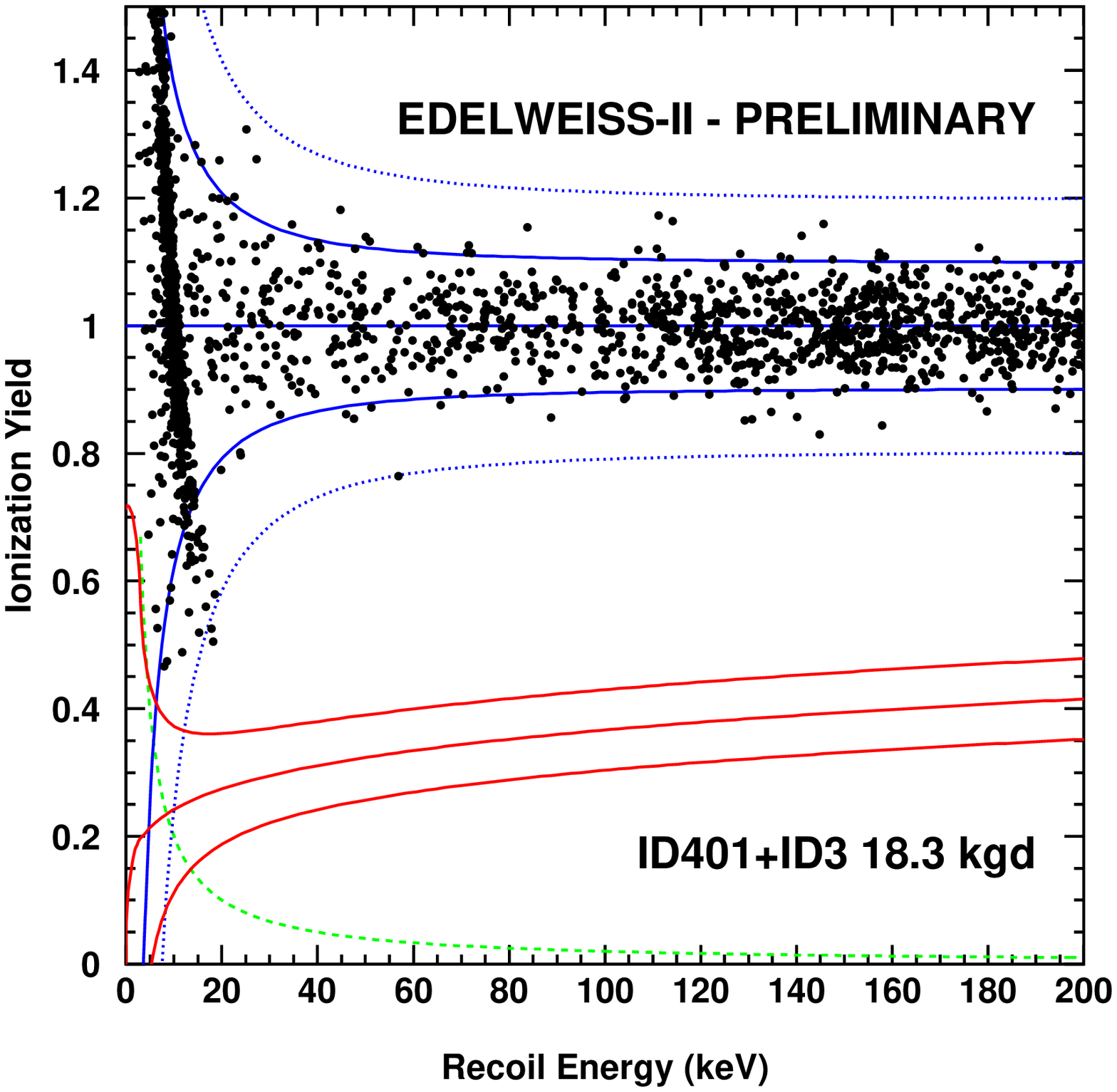}
\caption{\label{id3} Ionization yield vs recoil energy recorded in ID detectors for a fiducial exposure of 18.3~kg.d.}
\end{minipage}\hspace{4pc}%
\begin{minipage}{17pc}
\includegraphics[width=20pc]{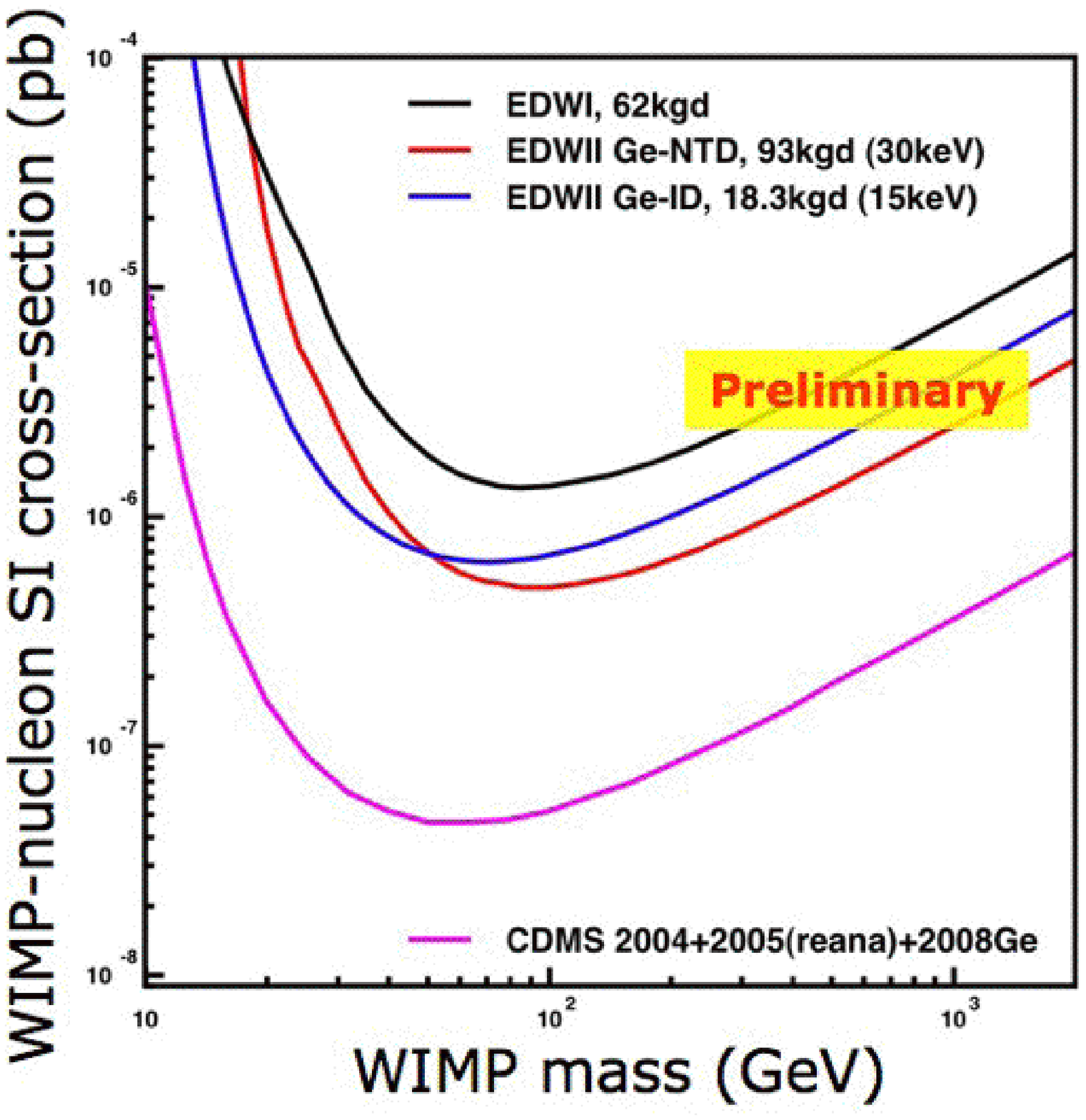}
\caption{\label{id4} 90$\%$ C.L. limits for spin-independent scattering cross-section for WIMPs as a function of WIMP mass.}
\end{minipage} 
\end{figure}

In 2008, a fiducial exposure of 18.3~kg.d was achieved using two 400 g ID detectors. Figure~\ref{id3} shows the ionization yield as a function of recoil energy for background runs and no events have been recorded in the nuclear recoil band down to a threshold of 10~keV in recoil energy. These data were interpreted in terms of limits for spin-independent scattering cross-section for WIMPs as a function of their mass, as shown in figure~\ref{id4}. 
This limit is comparable to the one obtained with the standard EDELWEISS detectors with an exposure of 93.5~kg.d despite the factor five in exposure. This demonstrates the importance of an active surface event rejection.\\
The EDELWEISS collaboration is presently operating ten 400 g ID detectors. A sensitivity of 4$\times$10$^{-8}$~pb is expected to be reached by 2010.  Further improvements are expected using new detectors with an increased fiducial volume. On a longer timescale, ID detectors are well-fitted for future larger scale experiments (100~kg to 1~ton) such as EURECA~\cite{eureca}. 

\section{Conlusion}
The EDELWEISS-II setup has been validated with both calibration and background runs. The results obtained with standard detectors show the importance of an active rejection of surface events. The new generation of ID detectors is also validated. With these new detectors showing an excellent $\beta$ rejection, EDELWEISS-II is now aiming at sensitivities below 10$^{-8}$~pb.


\section*{References}
\begin{thebibliography}{99}
\bibitem{cdms1} Z. Ahmed et al., {\it Phys. Rev. Lett.}, 2009 {\bf 102}, 011301
\bibitem{xenon} J. Angle et al., {\it Phys. Rev. Lett}, 2008 {\bf 100}, 021303
\bibitem{cresst} G. Angloher et al., {\it Astropart. Phys.}, 2009 {\bf 31}, 270 
\bibitem{neutron} G. Chardin and G. Gerbier, in Proceedings of th 4th International 
Workshop on Identification of Dark Matter (IDM2002), eds N.J. Spooner and V. Kudryavstev
\bibitem{simon} S. Fiorucci et al., {\it  Astropart. Phys.}, 2007 {\bf 28}, 143 
\bibitem{edel6} V. Sanglard et al., {\it Phys. Rev. D}, 2005 {\bf 71}, 122002
\bibitem{veto} K. Eitel, these proceedings
\bibitem{edel3} X.F. Navick et al., {\it Nucl. Inst. Meth. A}, 2000 {\bf 444}, 361
\bibitem{id1} X. Defay et al., in the Proc. of the 12$^{th}$ Intl. Workshop on Low Temperature (LTD12), Paris 2007, {\it J. Low Temp. Phys.}, 2008 {\bf 151}, 896
\bibitem{id2} A. Broniatowski et al., in the Proc. of the 12$^{th}$ Intl. Workshop on Low Temperature (LTD12), Paris 2007, {\it J. Low Temp. Phys.}, 2008 {\bf 151}, 830
\bibitem{id3} A. Broniatowski et al., accepted in  {\it  Phys. Lett. B} [arXiv:0905.0753]
\bibitem{eureca} H. Kraus et al.,  {\it  Nucl. Phys. B (Proc. Suppl.)}, 2007 {\bf 173}, 168
\end{thebibliography}
\end{document}